\documentclass[twocolumn,aps,prb,preprintnumbers,footinbib,floatfix,superscriptaddress]{revtex4}
\usepackage{amsmath}
\usepackage{amssymb}
\usepackage{color}
\usepackage{graphicx}

\newcommand{\be}{\begin{equation}}
\newcommand{\ee}{\end{equation}}

\newcommand{\lf}{\left}
\newcommand{\rg}{\right}

\newcommand{\bea}{\begin{eqnarray}}
\newcommand{\eea}{\end{eqnarray}}

\begin{document}

\title{\Large{Anisotropy of transport in bulk Rashba metals}}

\author{Valentina Brosco}
\affiliation{Istituto Officina dei Materiali (IOM) and Scuola Internazionale Superiore
di Studi Avanzati (SISSA), Via Bonomea 265, 34136 Trieste, Italy}
\author{Claudio Grimaldi}\affiliation{Laboratory of Physics of Complex Matter, Ecole Polytechnique F\'ed\'erale de
Lausanne, Station 3, CH-1015 Lausanne, Switzerland}

\begin{abstract}
The recent experimental discovery of three-dimensional (3D) materials hosting a strong Rashba spin-orbit coupling calls 
for the theoretical investigation of their transport properties.
Here we study the zero temperature dc conductivity of a 3D Rashba metal in the presence of static diluted impurities.
We show that, at variance with the two-dimensional case, in 3D systems spin-orbit coupling affects dc charge transport in all density regimes.
We find in particular that the effect of spin-orbit interaction strongly depends on the 
direction of the current, and we show that this yields strongly anisotropic transport characteristics.
In the dominant spin-orbit coupling regime where only the lowest band is occupied, the conductivity anisotropy is governed 
entirely by the anomalous component of the renormalized current. We propose that measurements of the conductivity anisotropy in bulk 
Rashba metals may give a direct experimental assessment of the spin-orbit strength.
\end{abstract}

\maketitle

\section{Introduction}
\label{intro}
Rashba spin-orbit (SO) interaction \cite{rashba1960,Bychkov1984} is commonly associated to two dimensional (2D) or quasi-2D
systems such as asymmetric  quantum well heterostructures, surface states and interfaces \cite{Manchon2015}.
It ultimately arises from the structural inversion asymmetry of these 2D systems that locks the spin and momentum degrees of freedom 
and lifts the spin degeneracy without the aid  of external magnetic fields. 
Besides its crucial relevance for spintronics \cite{awschalom2007,sinova2015} and spin-orbitronics \cite{kuschel2015} applications, 
Rashba SO interaction may have profound consequences both on the normal-state transport \cite{maekawa1981} and on 
superconductivity \cite{gorkov2001}. Moreover it may give rise to new topological phases of matter hosting Majorana fermions \cite{kitaev2001}  
or other topological  gapless states \cite{Hasan2010}. 

In the last decade a wealth of experiments on interfaces between complex oxides \cite{Shalom2010,Caviglia2010}, surface alloys \cite{Ast2007,Gierz2009} and polar semiconductors \cite{Ishizaka2011,Eremeev2012,Sakano2013} demonstrated that very large values of Rashba coupling  can be achieved going outside the realm of III-V semiconductor quantum wells, see {\sl e.g.} Refs. \cite{Ast2007,Gierz2009,Sakano2013} where Rashba energies, $E_0$, of the order of $100$ meV have been reported. 

Most importantly, the discovery of a giant Rashba effect in {\sl bulk} BiTe$X$ ($X$ = Br, Cl, or I) \cite{Lee2011,Ishizaka2011,Landolt2012,Martin2016}, in
GeTe \cite{DiSante2013,Liebmann2016,Krempasky2016}, and in organometal halide perovskites such as CH$_3$NH$_3$Pb$X_3$ ($X$ = I, Br) \cite{Niesner2016}, 
opened up new research pathways for solid-state physics and spintronics. 
This is illustrated, for example, by the connection between ferroelectricity and the Rashba SO interaction in GeTe \cite{DiSante2013}, or by the
the exceptionally long lifetime of photoelectrons in CH$_3$NH$_3$PbI$_3$ which is considered to be promoted by the giant
Rashba splitting of this semiconductor \cite{Zheng2015}. 
Although stoichiometric BiTe$X$, GeTe, and CH$_3$NH$_3$Pb$X_3$ are semiconductors with band gaps ranging
from $\approx 0.4$ to $\approx 2$ eV, natural defects, nonstoichiometry or doping can induce a finite population of electrons (holes)
in the conduction (valence) band, giving therefore rise to low density bulk Rashba metals with a finite Fermi energy $E_F$. 
In this respect, bismuth tellurohalides BiTeX offer a particularly interesting playground as their $E_F$ can be tuned to values
larger or smaller than the Rashba splitting $E_0$ \cite{Ishizaka2011,Lee2011,Landolt2012,Xiang2015,Martin2016}.

Contrary to their 2D analogues, in three dimensional (3D) Rashba metals, the Rashba SO coupling is an intrinsic effect of the  non-centrosymmetric 
crystal structure or of the site dipole field \cite{zhang2014} which gives rise to an effective low-energy Hamiltonian displaying a 
torus-shaped Fermi surface and characterized by a Rashba vector, below denoted $\mbox{\boldmath$\alpha$}$, pointing
along the inversion symmetry breaking direction.

In this paper we study the dc conductivity of a 3D Rashba metal in the presence of static disorder with electron densities
ranging from the high-density (HD) regime, $E_F>E_0$, to the dominant SO (DSO) regime $E_F<E_0$. 
We show that {\sl in both regimes},
(i) the Fermi surface topology affects the density dependence of the elastic quasi-particle scattering rates;
(ii)  the SO coupling strongly renormalizes the velocity  in the plane perpendicular to 
$\mbox{\boldmath$\alpha$}$, yielding deviations from standard  Drude transport laws;
(iii) due to the different electron velocities in the direction parallel or perpendicular to $\mbox{\boldmath$\alpha$}$, dc charge 
transport displays an anisotropy that can be directly related to Rashba spin-orbit coupling.

In particular, using Kubo linear response theory, we find that the conductivity $\sigma_\perp$ within the plane perpendicular to 
$\mbox{\boldmath$\alpha$}$ is strongly affected by the SO-induced  renormalization of the in-plane velocity and it shows an unconventional
behavior analogous to that predicted for the conductivity of 2D Rashba metals \cite{Brosco2016}. On the contrary, 
the velocity component along $\mbox{\boldmath$\alpha$}$ is not renormalized and the corresponding conductivity $\sigma_\parallel$
follows the Drude-like behavior expected for a metal with toroidal Fermi surface.   
In the DSO regime, $E_F<E_0$, we find that in the limit of weak disorder the SO-induced conductivity anisotropy is 
entirely dominated by the renormalization of the in-plane velocity and that it accurately follows 
\begin{equation}
\label{ani0}
\frac{\sigma_\perp}{\sigma_\parallel}=\frac{1}{2}\left(1+\frac{E_F}{2E_0}\right)\frac{m_\parallel}{m_\perp},
\end{equation}
where $m_\parallel$ and $m_\perp$ are the effective electron masses along and perpendicular to $\mbox{\boldmath$\alpha$}$, respectively.
We also show that, at variance with the 2D case \cite{Brosco2016}, sizable effects of the SO coupling on transport are not confined to the DSO regime.
Dc conductivity experiments can therefore directly measure the SO renormalization of the electron velocity and the Rashba energy 
splitting $E_0$,  and they thus represent a valuable tool for the characterization of bulk Rashba metals. 

The paper is organized as follows. After introducing the model, we briefly discuss the single-particle properties. We then come 
to the transport properties, explicitly calculating the renormalization of the current and the conductivity. 

\section{Model}
We consider the following low-energy Hamiltonian 
\begin{equation}
\label{H0}
H_0=\frac{\hbar^2k^2_\perp \sigma_0}{2m_\perp}+\frac{\hbar^2k^2_z \sigma_0}{2m_\parallel}+\hbar\, \alpha(\hat{z}\times\mathbf{k})\cdot\mbox{\boldmath$\sigma$},
\end{equation}
where $\mathbf{k}=(k_x,k_y,k_z)$ is the electron wave number, $\mbox{\boldmath$\alpha$}=\alpha\hat{z}$ is the Rashba vector pointing 
along the $z$-axis \cite{note}, and $k_\perp=\sqrt{k_x^2+k_y^2}$. We indicated as $\sigma_0$ the $2\times2$ identity matrix 
and as $\mbox{\boldmath$\sigma$}$ the spin-vector operator with components given by the Pauli matrices $\sigma_x$, $\sigma_y$, and $\sigma_z$. 

Let us remark that bulk Rahsba metals as those mentioned in the introduction may feature also a small Dresselhaus spin-orbit coupling. 
The latter may distort the bands as discussed, {\sl e.g.} in Refs. [\onlinecite{Ishizaka2011},\onlinecite{winkler}]. Here for the sake 
of simplicity, we focus only on Rashba coupling, as a matter of fact, the effects described in the present work can be ascribed to this 
type of coupling.  
\begin{figure}[t]
\begin{center}
\includegraphics[width=8.5cm,clip=true]{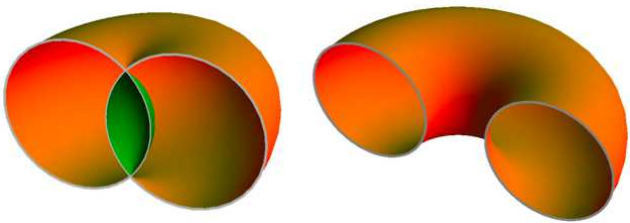}
  \caption{Topology of the Fermi surfaces of a bulk Rashba metal. For $E_F>E_0$ the Fermi surface is a spindle torus with inner and outer surfaces 
  corresponding to $s=+1$ (green) and $s=-1$ (red), respectively (left). In the DSO regime $E_F\leq E_0$, the Fermi level crosses only the 
  $s=-1$ band and the Fermi surface is a simple torus (right).}\label{fig1}
  \end{center}
\end{figure}

Diagonalization of Eq.~\eqref{H0} gives an electron dispersion consisting of two bands that apart from a constant energy shift
are given by
\begin{equation}
\label{e1}
E_{\mathbf{k}s}=\frac{\hbar^2}{2m_\perp}(k_\perp+s k_0)^2+\frac{\hbar^2k_z^2}{2m_\parallel}, 
\end{equation}
where $s=\pm 1$ is the helicity number and  $k_0=m_\perp\alpha/\hbar$ is the Rashba momentum. The SO splitting $E_0=\hbar^2 k_0^2/2m_\perp$,
which corresponds to the energy difference between the bottom of the $s=-1$ band and the degeneracy point at $k=0$, is a characteristic
energy scale of the system. In the HD regime the Fermi level crosses the bands of both helicities and the  Fermi surface
is a spindle torus, with the inner and outer sheets corresponding respectively to $s=+1$ and $s=-1$, as shown in the left panel of Fig.~\ref{fig1}. 
In this regime, the density of state (DOS) at the Fermi level is given by 
\be N(E_F)=a\sqrt{E_F-E_0}{\cal F}(\sqrt{E_0/(E_F-E_0)})\label{nef1}\ee
with  $a=\sqrt{2 m_\parallel}m_\perp/(\pi^2\hbar^3)$ and ${\cal F}(x)=1+ x\arctan(x)$.
Note that $N(E_F)$ reduces the well-known expression for the density of states of a 3D electron gas $N_0(E_F)=a\sqrt{E_F}$ asymptotically 
for $E_F\gg E_0$ \cite{Cappelluti2007}. In the DSO regime the Fermi level crosses only the $s=-1$ band and the Fermi surface becomes a 
ring torus (Fig.~\ref{fig1}, right panel) \cite{Cappelluti2007,Landolt2012}. 
In this regime the DOS is independent of $E_F$, as in 2D electron gases, and it is given by
\be
N(E_F)=a\pi\sqrt{E_0}/2.\label{nef2}\ee 
A similar effective reduction of dimensionality, 
but this time from 2D to 1D, is observed in 2D Rashba metals \cite{Cappelluti2007}.
A peculiarity of the two-dimensional case is that the effect of  the Rashba coupling on the DOS essentially disappear as soon as the system enters the HD regime {i.e.} in this regime  the contributions of the two bands sum up to give the standard result for the DOS of a 2D electron gas, $N(E_F)=m/\pi$.
From equation (\ref{nef1}),  we notice that instead in three dimensions the effects of Rashba spin-orbit coupling on the density of states are finite in both regimes.
This is due to the fact that, even for $E_F>E_0$,  the structure of the Fermi surface  at a given $k_z$ changes from an effectively 
high-density to an effectively dominant spin-orbit regime as $k_z$ increases. Both situations are thus found  and contribute to the 
different integrals.

As one may expect, the  dependence of $N(E_F)$ on $E_0$ affects the momentum relaxation rate of the carriers and consequently 
their transport properties. An additional effect of the Rashba SO coupling on transport stems however also from the 
renormalization of the electron velocity $\mathbf{v}$.
In the clean limit, the latter corresponds to the following operator $\mathbf{v}_i(\mathbf{k})=(e/\hbar)\partial H_0/\partial k_i$, {\sl i.e.} 
\begin{align}
\label{currx}
\mathbf{v}_{x(y)}(\mathbf{k})&=\frac{\hbar k_{x(y)}}{m_\perp}+(-)\alpha \sigma_{y(x)}, \\
\label{currz}
\mathbf{v}_z(\mathbf{k})&=\frac{\hbar k_z}{m_\parallel}.
\end{align}
From the above equations we see that only the components of the velocity in the $(x,y)$-plane, orthogonal to $\mbox{\boldmath$\alpha$}$, acquire 
an anomalous spin-dependent term while the $z$-component is not modified by the Rashba interaction.
This SO-induced direction dependence of the bare velocity survives in the presence of disorder and it is ultimately responsible for 
the anisotropy of the dc conductivity, as we now show.

\section{Self-consistent Born transport equations}
The low-temperature conductivity is typically limited by the scattering of the electrons by defects and impurities yielding, in the simplest approximation, white noise static disorder with gaussian correlations, 
$\langle V_\textrm{imp}(\mathbf{r})V_\textrm{imp}(\mathbf{r}')\rangle_\textrm{imp}=n_i v_\textrm{imp}^2\delta(\mathbf{r}-\mathbf{r}')$,
where $n_i$ is the density of the scattering centers and $v_\textrm{imp}$ is the scattering strength.
At zero temperature, the Kubo formula for the conductivity along the direction $i$ of a 3D Rashba metal is \cite{mahan}
\begin{equation}
\label{sigma1}
\sigma_{ii}=\frac{\hbar}{2\pi}[P_{ii}^{AR}-\textrm{Re}(P_{ii}^{RR})],
\end{equation}
where $P_{ii}^{AR}$ and $P_{ii}^{RR}$ are respectively the advanced-retarded and the retarded-retarded
current-current correlation functions given by ($L,M=A,R$):
\begin{equation}
\label{P1}
P_{ii}^{LM}=\int\!\frac{d\mathbf{k}}{(2\pi)^3}\textrm{Tr}\left[\mathbf{j}_i(\mathbf{k})\mathbf{G}^L(\mathbf{k})
\mathbf{J}_i^{LM}(\mathbf{k})\mathbf{G}^M(\mathbf{k})\right].
\end{equation} 
In the above expression, $\mathbf{j}_i(\mathbf{k})=e\,\mathbf{v}_i(\mathbf{k})$ is the bare charge current, 
$\mathbf{G}^L(\mathbf{k})$ with $L=R(A)$ denotes the zero-frequency retarded (advanced) electronic Green's function 
and $\mathbf{J}^{LM}_i(\mathbf{k})$ is the dressed operator for the current along $i$. 
Within the self-consistent Born approximation, the retarded Green's function can be cast as 
\begin{equation}
\label{green1}
\mathbf{G}^{R(A)}(\mathbf{k})
=\frac{1}{2}\sum_s g_s^{R(A)}(\mathbf{k})\left[1+s(\hat{z}\times\hat{k})\cdot\mbox{\boldmath$\sigma$}\right],
\end{equation}
where $g_s^{R(A)}(\mathbf{k})=\left[E_F-E_{\mathbf{k}s}-\Sigma^{R(A)}\right]^{-1}$
is the electron propagator in the helicity basis and $\Sigma^{R(A)}$
is the retarded (advanced) electron self-energy which, within our assumptions, is spin-independent:
\begin{equation}
\label{self1}
\Sigma^{R(A)}=\frac{n_i v_\textrm{imp}^2}{2}\int\!\frac{d\mathbf{k}}{(2\pi)^3}\sum_sg_s^{R(A)}(\mathbf{k}).
\end{equation}
The dressed current operator is the solution of the following ladder equation:
\begin{equation}
\label{j2}
\mathbf{J}_i^{LM}(\mathbf{k})=\mathbf{j}_i(\mathbf{k})+n_i v_\textrm{imp}^2
\int\!\frac{d\mathbf{k}'}{(2\pi)^3}\mathbf{G}^L(\mathbf{k}')\mathbf{J}_i^{LM}(\mathbf{k}')\mathbf{G}^M(\mathbf{k}').
\end{equation}

Let us focus for the moment on the conductivity in the $(x,y)$-plane orthogonal to the
Rashba vector. Without loss of generality, we set $i=x$ in Eqs.~\eqref{sigma1}-\eqref{j2} and
identify $\sigma_\perp$ with $\sigma_{xx}$. Using Eqs.~\eqref{currx} and \eqref{j2}, it can be easily shown that the dressed current has the same structure of the bare current but with a renormalized value of $\alpha$, {\sl i.e.} we can write 
\be \mathbf{J}_x^{LM}=e\hbar k_x/m_\perp+e \sigma_y\tilde \alpha^{LM},\ee
where $\tilde \alpha^{LM}$ denotes the renormalized anomalous velocity and it is given by the following equation:
\begin{equation}
\label{lambda2}
\tilde\alpha^{LM}=\frac{\alpha+\alpha_0^{LM}}{1-A^{LM}},
\end{equation}
where 
$\alpha_0^{LM}$ and $A^{LM}$ are defined by the following equations
\begin{align} 
\alpha_0^{LM}&=\frac{n_iv_{\rm imp}^2\hbar}{4m_\perp}\!\int\!\frac{d\mathbf{k}}{(2\pi)^3}\!\sum_ssg_s^L(\mathbf{k})g_s^M(\mathbf{k})k_\perp,\label{alpha0}\\
\label{A}
A^{LM}&=\frac{n_iv_{\rm imp}^2}{4} \int\!\frac{d\mathbf{k}}{(2\pi)^3}\sum_{s,s'}g_s^L(\mathbf{k})g_{s'}^M(\mathbf{k}).
\end{align}

By using Eqs.~\eqref{green1} and \eqref{j2}, after some algebra, it can be shown that the current-current correlation function is given by
\begin{equation}
\label{Pxx1}
P_{xx}^{LM}=e^2\bigg[P_0^{LM}+\frac{2(1-A^{LM})}{n_i v_{\rm imp}^2} \tilde \alpha^{LM}\left(\tilde \alpha^{LM}-\alpha\right)\bigg],
\end{equation}
where we defined $P_0^{LM}$ as 
\begin{equation}
\label{P01}
P_0^{LM}=\frac{\hbar^2}{2m_\perp^2}\int\!\frac{d\mathbf{k}}{(2\pi)^3}\sum_s g_s^L(\mathbf{k})g_s^M(\mathbf{k})(k_\perp^2+s k_\perp k_0).\\
\end{equation}

Equations  (\ref{sigma1}) and (\ref{Pxx1}) along with the self-consistent self-energy and vertex equations, Eqs.\eqref{self1} and \eqref{lambda2},
and with the definitions of $A$, $\alpha_0$, and $P_0$ given in Eqs.(\ref{alpha0}), (\ref{A}) and (\ref{P01}) will be used  below for a fully self-consistent, 
numerical calculation of the conductivity $\sigma_\perp$.
Although these equations have exactly the same structure as those found for the 2D Rashba model \cite{Brosco2016},
their application to the three-dimensional case yields a little surprise: in 3D the renormalized anomalous velocity, $\tilde \alpha^{AR}$, does not renormalizes 
to zero for $E_F>E_0$, as instead it happens in standard 2D Rashba metals \cite{Brosco2016}. 
To understand this point and its consequences on dc-transport, we now consider the weak-disorder limit of our model and we  estimate the different contributions to the  conductivity $\sigma_\perp$  analytically.

\subsection{Analytical estimate of the anomalous velocity and of the conductivity}

In the weak disorder limit, we can neglect the $RR$ contribution to Eq. \eqref{sigma1} and approximate the spectral function as
\be g_s^A(\mathbf{k})g_s^R(\mathbf{k})\simeq (\pi/\Gamma)\delta\!(E_{\mathbf{k}s}-E_F)\label{WDL}\ee  with $\Gamma$ denoting
the elastic scattering rate, 
\begin{equation} 
\Gamma=i\Sigma^{R}=\pi n_iv_\textrm{imp}^2N(E_F)/2.
\end{equation}
Using the self-consistent self-energy equation it is then straightforward to show that $A^{AR}=1/2$. 
Furthermore, by inserting Eq.\eqref{WDL} in the definition of $P_0^{AR}$ and integrating by parts with respect to $k_{\perp}$, one can easily prove that
(see Appendix) 
\begin{equation}
\label{P02}  
P_0^{AR}=\frac{2\pi}{\hbar}\frac{\tau(n)n}{m_\perp},
\end{equation}
where $\tau(n)=\hbar/2\Gamma(n)$  and $n$ denotes the electron density, $n=\int_0^{E_F} \!\! N(E)dE$.
For $E_F>E_0$, the latter is given by
\begin{equation}
\label{density}
n= a\sqrt{E_F-E_0}\bigg[\frac{E_0-E_F}{3}+E_F {\cal F}\big(\sqrt{\frac{E_0}{E_F-E_0}}\bigg)\bigg], 
\end{equation}
while for $E_F\leq E_0$ it becomes
\be n=a\frac{\pi}{2}\sqrt{E_0}E_F. \label{densitylow}\ee
Here we notice once again the dimensional cross-over in the dependence of the  density on $E_F$ as the system goes from the HD to the DSO regime.
Eventually, an explicit calculation of the integral $\alpha_0^{AR}$, Eq.~\eqref{alpha0}, yields  the following expressions for the renormalized vertex $\tilde \alpha^{AR}$:
\be \label{tildealpha1}
\tilde \alpha^{AR}=\frac{\alpha}{2}\lf[1-\sqrt{\frac{{E_F-E_0}}{E_0}}\frac{\arctan{\sqrt{E_0/(E_F-E_0)}}}{{\cal F}(\sqrt{E_0/(E_F-E_0)})}\rg] 
\ee
for $E_F>E_0$,
while
\begin{equation}\label{tildealpha2}
\tilde \alpha^{AR}=\alpha\lf(1-\frac{E_F}{2 E_0}\rg)
\ee
 for $E_F<E_0$.
From Eq.~\eqref{tildealpha1} we see explicitly that in three dimensions the anomalous current is finite both in the DSO and, at variance 
with the two-dimensional case, also in the HD regime $E_F>E_0$, while it vanishes only for $E_F \gg E_0$. In this limit we find 
$\tilde\alpha^{AR}\propto E_0/E_F$ which in terms of the density reduces to $\tilde\alpha^{AR}\propto (n_0/n)^{2/3}$, where $n_0=a\pi E_0^{3/2}/2$ is 
the density corresponding to $E_F=E_0$. 

Let us recall that the anomalous dressed velocity $\tilde\alpha^{LM}$, Eq.~\eqref{lambda2}, governs also the intrinsic spin-Hall effect in 
Rashba systems, which is the generation of a spin-polarized current transverse to the direction of an applied electric field 
(for a recent review see Ref.~\onlinecite{sinova2015}). 
In disordered 2D Rashba gases with $E_F>E_0$ the dc spin-Hall conductivity $\sigma^\textrm{sH}_{xy}$ is zero because $\tilde\alpha^{AR}=0$, while for 
$E_F<E_0$ the vanishing of $\sigma^\textrm{sH}_{xy}$ stems from a non-trivial cancellation between the $AR$ and $RR$ terms of the spin-Hall 
kernel \cite{Cappelluti2006}.   
In more general terms, the vanishing of the spin-Hall effect in Rashba systems with parabolic dispersion is the consequence of the vanishing of the
magnetization rate in the steady state \cite{Chalaev2005,Dimitrova2005}. 
Since this statement holds true also for 3D Rashba gases, the observation that $\tilde\alpha^{AR}\neq 0$ 
does not imply that there is in this case a finite dc spin-Hall response.

Let us now consider the conductivity. As one can infer from Eq.~\eqref{P02}, $P^{AR}_0$ yields a Drude-like contribution to the conductivity.
In the high density limit $n\gg n_0$ ($E_F\gg E_0$), due to the vanishing of $\tilde \alpha^{AR}$, this is the only contribution
and  we recover a Drude-like formula for the in-plane conductivity:
\begin{equation}
\label{sigmaxxgg}
\sigma_\perp\simeq \frac{e^2\tau(n)n}{m_\perp},\,\,\,\,\, n\gg n_0.
\end{equation}
Since $\tau\propto n^{-1/3}$ in this
limit, we obtain that $\sigma_\perp\propto n^{2/3}$, as in standard 3D electron gases. 
Note how equations \eqref{Pxx1} and \eqref{P02} demonstrate that
the recovering of a Drude-like behavior for the Rashba model is by itself a non-trivial result, accessible only if the renormalized anomalous current vanishes.
As the density becomes comparable with $n_0$, the anomalous velocity \eqref{tildealpha1} increases and, the conductivity starts to deviate from the Dude theory. Eventually, in
the DSO regime  $n\leq n_0$ ($E_F\leq E_0$) we find
\begin{equation}
\label{sigmaxxll}
\sigma_\perp=\frac{e^2\tau_0 n}{m_\perp}\left(1-\frac{\tilde\alpha^{AR}}{2\alpha}\right)=
\frac{e^2\tau_0 n}{2m_\perp}\left(1+\frac{n}{2 n_0}\right),\,\,\,\,\, n\leq n_0
\end{equation}
where the elastic scattering time, $\tau_0$, is independent of the density because the DOS is constant for $E_F\leq E_0$.
The above expression markedly deviates from the 3D Drude-like behavior of Eq.~\eqref{sigmaxxgg}. 
In particular, the low density limit $n\ll n_0$ of Eq.~\eqref{sigmaxxll} scales as $\sigma_\perp\propto n$, as the Drude
conductivity of an effectively 2D electron gas.

We now turn to evaluate the conductivity along the direction $\hat{z}$ of the Rashba field: $\sigma_\parallel=\sigma_{zz}$.
Since the current operator $\mathbf{j}_z(\mathbf{k})$ in Eq.~\eqref{currz} is diagonal in spin-space, there is no renormalization
due to the Rashba interaction and the solution of the ladder equation
\eqref{j2} for the current is simply $\mathbf{J}_z(\mathbf{k})=\mathbf{j}_z(\mathbf{k})$.
The current-current correlation function thus reduces to
\begin{equation}
\label{Pz2}
P_{zz}^{LM}=\left(\frac{e\hbar}{m_\parallel}\right)^2\sum_s\int\!\frac{d\mathbf{k}}{(2\pi)^3}k_z^2g_s^L(\mathbf{k})g_s^M(\mathbf{k}).
\end{equation}
In the limit of weak disorder,  it is straightforward to show that $P_{zz}^{AR}$ is 
proportional to $n/\Gamma$ (see the Appendix), where the electron density $n$ is given in Eqs.~\eqref{density} and \eqref{densitylow}, and 
$\sigma_\parallel=\hbar P_{zz}^{AR}/(2\pi)$ becomes
\begin{equation}
\label{sigmaz1}
\sigma_\parallel=\frac{e^2\tau(n)n}{m_\parallel},
\end{equation}
from which we recover the 3D Drude scaling $\sigma_\parallel\propto n^{2/3}$ for $n\gg n_0$ and the 2D Drude behavior $\sigma_\parallel\propto n$
for $n\leq n_0$.
\begin{figure}[t]
\begin{center}
\includegraphics[width=8.5cm,clip=true]{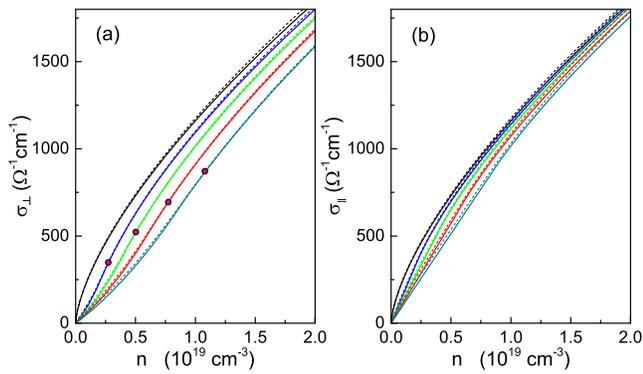}
  \caption{(a) Direct current conductivity $\sigma_\perp$ within the plane orthogonal to the direction of the Rashba vector as a 
	function of the electron density $n$ for $E_0=0,\, 40,\,60,\,80,$ and $100$ meV (from top to bottom), $m_\perp=m_\parallel=0.1 m_e$,
	and the scattering rate is set equal to $\Gamma=10$ meV for $E_0=0$ and $n=2\times 10 ^{19}$ cm$^{-3}$. The solid lines 
	denote the analytic results for the weak disorder limit, while the dashed lines are the corresponding conductivity
	values obtained from the numerical self-consistent calculation. The circles mark the values of $\sigma_\perp$ corresponding
	to $n=n_0=a\pi E_0^{3/2}/2$. (b) Direct current conductivity $\sigma_\parallel$ along the direction
	of the Rashba vector for the same parameters of (a).}\label{fig2}
  \end{center}
\end{figure}

\section{Numerical results and transport anisotropy}
Our analytic results of $\sigma_\perp$ are compared with the numerical self-consistent calculations of the conductivity in Fig.~\ref{fig2}(a).
We have performed the
calculations by setting $m_\perp=m_\parallel=0.1m_e$, where $m_e$ is the free-electron mass, and by requiring that $\Gamma=10$ meV
for $n=2\times 10^{19}$ cm$^{-3}$ and $E_0=0$. The numerical results (dashed lines) are almost indistinguishable from
the analytic $\sigma_\perp$ (solid lines), as shown in Fig.~\ref{fig2}(a) for values of the Rashba splitting ranging from $E_0=0$ 
to $E_0=100$ meV (from top to bottom). The values of $\sigma_\perp$ corresponding to $n=n_0$ are marked
by the filled circles.

A comparison between the analytical and numerical results for the conductivity parallel to Rashba vector is shown
in Fig.~\ref{fig2}(b) for the same parameters used in Fig.~\ref{fig2}(a). The self-consistent solution for $\sigma_\parallel$ 
(dashed lines) shows excellent agreement with the analytic expression, Eq.\eqref{sigmaz1} represented by the solid lines. 

\begin{figure}[t]
\begin{center}
\includegraphics[width=8.2cm,clip=true]{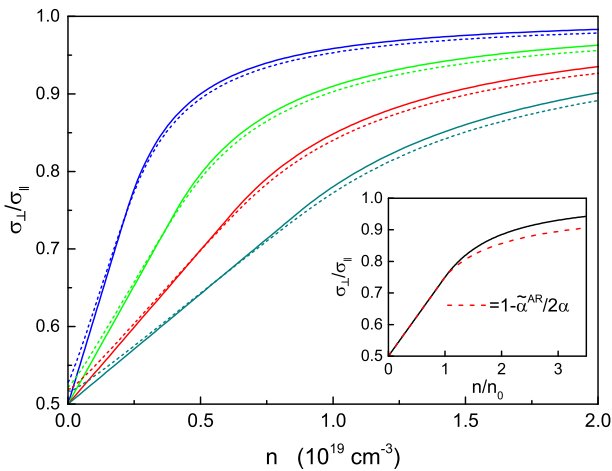}
  \caption{Direct current conductivity anisotropy $\sigma_\perp/\sigma_\parallel$ for the same parameters of Fig.~\ref{fig2}. Inset: the weak disorder
	limit of $\sigma_\perp/\sigma_\parallel$ as a function of the reduced density $n/n_0$. The dashed line denotes $1-\tilde\alpha^{AR}/2\alpha$,
	where $\tilde\alpha^{AR}$ is the SO renormalized the electron velocity.}\label{fig3}
  \end{center}
\end{figure}

We see from Figs.~\ref{fig2}(a) and \ref{fig2}(b) that $\sigma_\parallel$ coincides with
$\sigma_\perp$ only for $E_0=0$ (if $m_\perp=m_\parallel$), while for nonzero values of the Rashba splitting $\sigma_\parallel>\sigma_\perp$ 
because the SO interaction affects the current only for electron velocities along directions orthogonal to the Rashba field. 
The resulting conductivity anisotropy $\sigma_\perp/\sigma_\parallel$ in the limit of weak disorder is shown in Fig.~\ref{fig3} by solid lines.
In the DSO regime, $\sigma_\perp/\sigma_\parallel$ is expressed analytically from Eqs.~\eqref{sigmaxxll} and \eqref{sigmaz1}:
\begin{equation}
\label{ani1}
\frac{\sigma_\perp}{\sigma_\parallel}=\left(1-\frac{\tilde\alpha^{AR}}{2\alpha}\right)\frac{m_\parallel}{m_\perp}
=\frac{1}{2}\left(1+\frac{n}{2n_0}\right)\frac{m_\parallel}{m_\perp},\,\,\,n\leq n_0
\end{equation}
from which we recover Eq.~\eqref{ani0} since $n/n_0=E_F/E_0$ in this regime. Equation \eqref{ani1}
implies therefore that a measurement of the dc conductivity anisotropy  in the dominant SO regime can 
give direct experimental access to the anomalous vertex $\tilde\alpha^{AR}$ and to the Rashba splitting.
Comparison with the self-consistent results (dashed lines) shows the accuracy of Eq.~\eqref{ani1} for $m_\parallel=m_\perp$. For
$n>n_0$, the conductivity anisotropy starts to deviate from $1-\tilde\alpha^{AR}/2\alpha$, as shown in the inset of
Fig.~\ref{fig3}. In this regime, however, $\sigma_\perp/\sigma_\parallel$ as obtained from our analytical results
can still be used to estimate $n_0$ (and so $E_0$) from measurements of the conductivity anisotropy induced by the Rashba interaction.
Finally, for $n\gg n_0$ ($E_F\gg E_0$) the effects of the SO coupling becomes negligible and the conductivity anisotropy
is governed solely by the difference of the effective masses $m_\perp$ and $m_\parallel$: 
$\sigma_\perp/\sigma_\parallel\rightarrow m_\parallel /m_\perp$.

\section{conclusions}
We conclude by briefly discussing how the SO-induced anisotropy of transport here predicted can be observed in real 
bulk Rashba metals. 
As mentioned in the introduction, currently known semiconductors with strong Rashba interaction can be turned into
bulk Rashba metals by suitable doping or defect control. This is realized for example in bismuth tellurohalides where
$E_F$ can be tuned to values larger or smaller than $E_0$ ($\sim 100$ meV) in a way consistent with a rigid band shift \cite{Lee2011}.
The behavior of $\sigma_\perp/\sigma_\parallel$ illustrated in Fig.~\ref{fig3} could therefore be experimentally observed
by low-temperature dc measurements of $\sigma_\perp$ and $\sigma_\parallel$ as a function of the electron density $n$ extracted by,
e.g., Hall measurements. If the values of $n$ are such that the system is always in DSO regime, the linear relation of Eq.~\eqref{ani1}
would allow to extract the mass anisotropy $m_\parallel/m_\perp$ directly from the intercept of $\sigma_\perp/\sigma_\parallel$ at $n\rightarrow 0$
and the Rashba density $n_0$ from the slope of Eq.~\eqref{ani1}. More generally, the values of $n$ may range from
$n>n_0$ to $n<n_0$ (see for example Ref.~\onlinecite{Lee2011}). In this case, the two unknown parameters, $m_\perp/m_\parallel$ 
and $n_0$, can be extracted from a non-linear fit to our analytic formulas for  $\sigma_\perp/\sigma_\parallel$. Of course, these formulas
are valid as long as the electronic dispersion is parabolic (which implies that the density must be sufficiently small) and the 
Dresselhaus SO interaction can be neglected. Finally, we note that if $m_\parallel/m_\perp$ is known (for example from first principle calculations) 
the measurement of the transport anisotropy for a single value of $n$ suffices, in principle, to estimate $n_0$.

We thank L. Benfatto, E. Cappelluti, and R. Raimondi for stimulating discussions.
V.B. acknowledges financial support from MIUR through the PRIN 2015 program (Prot. 2015C5SEJJ001) and SISSA/CNR 
project "Superconductivity, Ferroelectricity and Magnetism in bad metals" (Prot. 232/2015)
 
\appendix
\section{Derivation of $P_0^{AR}$ and $P_{zz}^{AR}$}
\label{app}

Using the weak coupling approximation of Eq.~\eqref{WDL}, the $AR$ contribution of Eq.~\eqref{P01} reduces to:
\begin{equation}
\label{app1}
P_0^{AR}=\frac{\pi\hbar^2}{2m_\perp^2\Gamma}\int\!\frac{d\mathbf{k}}{(2\pi)^3}\sum_s \delta(E_{\mathbf{k}s}-E_F)(k_\perp^2+s k_\perp k_0),
\end{equation}
where $E_{\mathbf{k}s}$ is the electron dispersion given in Eq.~\eqref{e1}.
Noting that 
\begin{equation}
\label{app2}
\delta(E_{\mathbf{k}s}-E_F)(k_\perp^2+s k_\perp k_0)=-\frac{m_\perp k_\perp}{\hbar^2}\frac{\partial}{\partial k_\perp}
\theta(E_F-E_{\mathbf{k}s}),
\end{equation}
equation~\eqref{app1} can be rewritten in cylindrical coordinates for the wavevector as
\begin{equation}
\label{app3}
P_0^{AR}=-\frac{\pi}{2m_\perp\Gamma}\sum_s \iint\!\frac{dk_z dk_\perp k_\perp^2}{(2\pi)^2} \frac{\partial}{\partial k_\perp}
\theta(E_F-E_{\mathbf{k}s}).
\end{equation} 
Integration by parts with respect to $k_\perp$ gives 
\begin{equation}
\label{app4}
P_0^{AR}=\frac{\pi}{m_\perp\Gamma}\sum_s\iint\!\frac{dk_z dk_\perp k_\perp}{(2\pi)^2}\theta(E_F-E_{\mathbf{k}s})=\frac{\pi n}{m_\perp\Gamma},
\end{equation}
where $n=\sum_s\int d\mathbf{k}/(2\pi)^3\theta(E_F-E_{\mathbf{k}s})$ is the electron density. Using $\tau=\hbar/(2\Gamma)$ we recover
therefore Eq.~\eqref{P02}.

The calculation of $P_{zz}^{AR}$, Eq.~\eqref{Pz2}, follows similar steps. In the limit of weak disorder, the integrand of
Eq.~\eqref{Pz2} is rewritten as 
\begin{equation}
\label{app5}
\frac{\pi k_z ^2}{\Gamma}\delta(E_{\mathbf{k}s}-E_F)=-\frac{\pi k_z m_\parallel}{\Gamma\hbar^2}\frac{\partial}{\partial k_z} \theta(E_F-E_{\mathbf{k}s}).
\end{equation}
The integration by parts performed with respect to $k_z$ and the sum over $s$ give $n$, so that $P_0^{AR}=e^2\pi n/m_\parallel\Gamma$.


\begin{thebibliography}{70}
\bibitem{note} We have chosen the units such that $\alpha$ has dimension of a velocity.
\end{thebibliography}
\end{document}